\documentclass[prl,aps,pacs,twocolumn]{revtex4}
\usepackage{epsfig}
\usepackage{bm}
\usepackage{color}
\newcommand{\B}[1]{{\bm{#1}}}

\usepackage[latin1]{inputenc}
\newcommand{\beq}{\begin{equation}}
\newcommand{\eeq}{\end{equation}}
\newcommand{\bea}{\begin{eqnarray}}
\newcommand{\eea}{\end{eqnarray}}

\begin{document}
\title{Statistical Mechanics of the Glass Transition in One-Component Liquids with Anisotropic Potential}
\author{Valery Ilyin, Edan Lerner, Ting-Shek Lo and Itamar Procaccia }
\affiliation{Department of Chemical Physics, The Weizmann
Institute of Science, Rehovot 76100, Israel }
\begin{abstract}
 We study a recently introduced model of one-component glass-forming liquids whose constituents
 interact with anisotropic potential. This system is interesting per-se and as a model of liquids like
 glycerol (interacting via hydrogen bonds) which are excellent glass formers. We work out the statistical mechanics of this system, encoding the liquid and glass disorder using appropriate quasi-particles (36 of them).  The theory provides a full explanation of the glass transition phenomenology, including the identification of a diverging length scale and a relation between the structural changes and the diverging relaxation times.  
  \end{abstract}
\maketitle

The study of associated liquids like glycerol as glass formers has a long and rich history \cite{03RHGF}, but until now the role of the anisotropic hydrogen bonds, while clearly important in frustrating crystallization, has not been made explicit. Recently a model of one component liquids with anisotropic interaction potential was introduced \cite{06ST}, together with numerical simulations in two-dimensions that demonstrated clearly the importance of the anisotropic interaction in frustrating crystallization and allowing the formation of a glassy state of matter. This model is important in stressing the fact that even simple one-component liquids may not crystallize if the local symmetry of the interaction potential does not perfectly match the symmetry of the equilibrium crystal. It is worthwhile therefore to analyze further this example of glass formation and put it in the general context of the glass transition.  In this Letter we present a theory of this model, constructing its statistical mechanics and providing an understanding of the phenomenology of its glass transition, including an identification of a diverging length and explaining the diverging time scales. Our analysis allows putting this 
interesting example of glass formation on the same footing as other classical glass formers such as
binary mixtures with central potentials \cite{89DAY,99PH}, stressing the generality of the approach\cite{06ABIMPS,07HIMPS} and of the glass transition phenomenon at the same time. 

Particles of mass $m$ in this model  interact via 
\begin{equation}
U(r_{ij}, \theta_i,\theta_j) = \overline{U}(r_{ij}) +\Delta U(r_{ij}, \theta_i,\theta_j) \ , \label{pot}
\end{equation}
where $r_{ij}$ is the distance between the two particles $i$ and $j$. The first term on the RHS of (\ref{pot}) is the standard isotropic
Lennard-Jones potential
\begin{equation}
\overline{U}_{ij} =4 \epsilon\left[ \left(\frac{\sigma}{r_{ij}}\right)^{12}- \left(\frac{\sigma}{r_{ij}}\right)^{6}\right]\ ,  \label{LJ}
\end{equation}
whereas the anisotropic part of the potential is given by
\begin{eqnarray}
&&\Delta U(r_{ij}, \theta_i,\theta_j)=-4\epsilon \Delta \left(\frac{\sigma}{r_{ij}}\right)^{6}\Big[h\left(\frac{\theta_i-\theta_0}{\theta_c}\right)\nonumber\\&&+h\left(\frac{\theta_j-\theta_0}{\theta_c}\right)-\frac{64}{35\pi}\theta_c\Big] \  , \label{anisopot}\\
&&h(x)=(1-x^2)^3~{\rm for}~|x|<1\ ; h(x) =0\,{\rm for}\,|x|\ge 1 \ . \nonumber
\end{eqnarray}
Here $\theta_i$ ($\theta_j$) is the included angle between the relative vector $\B r_{ij}\equiv \B r_i-\B r_j$ and a unit vector $\B u_i$ ($\B u_j$) (referred to below as `spin') which represents the orientation of the axis of particle $i$ ($j$). The function $h((\theta-\theta_0)/\theta_c)$ (with $\theta_0=126^{o}$ and $\theta_c=53.1^{o}$) has a maximum at $\theta=\theta_0$, and thus $\theta_0$ is a favored value of $\theta_i$. Thus the anisotropic term in the potential favors structures of five-fold symmetry. The parameter $\Delta$ controls the tendency of five-fold symmetry, and therefore of the frustration against crystallization. The units of mass, length, time and temperature are $m$, $\sigma$, $\tau=\sigma\sqrt{m/\epsilon}$ and $\epsilon/k_B$, respectively, with $k_B$ being Boltzmann's constant. 

According to the numerical simulations presented in \cite{06ST}, for $\Delta<0.6$ this system crystallizes upon reducing the temperature. The ground state crystal has an elongated hexagonal structure with
anti-ferromagnetic ordering of the spins $\B u_i$, but the actual crystal that is obtained upon cooling is a ``plastic crystal'' with hexagonal spatial order but with spin disorder. For $\Delta>0.6$ the system fails to crystallize upon cooling. The relaxation times were measured by monitoring the rotational autocorrelation functions $C_R(t)\equiv (1/N)\sum_i \langle  \B u_i(t)\cdot \B u_i(0)\rangle$ which was
fitted to a stretched exponential form $C_R(t)\propto \exp[-(t/\tau_\alpha)^\beta]$. For $\Delta=0.6$ the relaxation is of Arrhenius form with a constant value $\beta\approx 0.95$ for $T>T_m=0.46$, but $\beta$ was fit separately for every temperature  $T<T_m$ where it decreases with temperature. The relaxation times were fitted to a Vogel-Fulcher law $\tau_\alpha=\tau_0 \exp[DT_0/(T-T_0)]$ which involves fitting the three free parameters $\tau_0=0.61$, $D=7.4$ and $T_0=0.099$ (in addition to $\beta$). 
We repeated the simulations of this model using Monte-Carlo methods in N-P-T ensemble
\cite{91AT}, finding results in agreement with the MD simulations
of \cite{06ST} in the same ensemble.

\begin{figure}
\hskip -1.0 cm
\epsfig{width=.47\textwidth,file=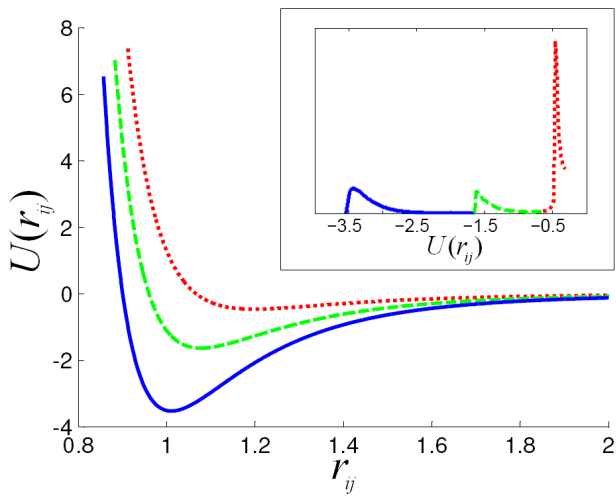}
\caption{(Color online). Potential curves for particle-pairs with two spins, one spin or no spin in favored
position (blue continuous line, green dashed or red dotted line respectively). Inset: the measured energies of particle pairs, falling in three distinct ranges with gaps between them, allowing us to define the quasi-particles. The peak in each colored curve corresponds to  the minimum in the main figure.}
\label{potentials}
\end{figure}

To construct the statistical mechanics of this system we recognize that the potential energy
between any pair of particles depends on their spin orientations. In Fig. \ref{potentials}
we present the three potentials between two particles, depending on the orientation of their spins relative to the inter-particle vector distance: lowest in energy  (in blue continuous line) is the case for which both have a favored spin orientation; middle, in green dashed line (high, in red dotted line) is the potential when one (none) of the spins are in a favored orientation. One sees that the minima of these potentials occur with significant gaps in their energies, 
allowing us to now measure the {\em average} energy of {\em pairs} of particles as a function of temperature. These averages fall in three distinct ranges, such that the range of variation of each energy is much smaller than the gaps between the energies, see inset in Fig. \ref{potentials}. This allows us to proceed to define quasi-species. We denote the three effective energies
below as $2E_b$ $2E_g$ and $2E_r$ respectively. Note that the spin orientations involved in each such mean energy
can fluctuate within a temperature-dependent range of angles. For the temperature range of interest the range of angles is in a sector of about $60^o$, but as this range of angles determines the
degeneracies that enter the statistical mechanics below,  we need to reconsider it carefully as explained in the sequel.
\begin{figure}
\hskip -1.25 cm
\epsfig{width=.30\textwidth,file=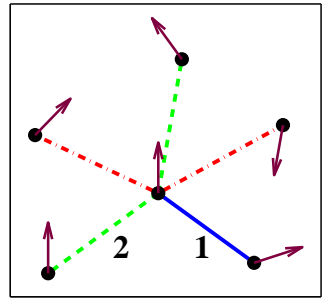}
\caption{(Color online). An example of an $n$-star with $n=5$,$i=2$, $j=2$ and $k=1$. The central particle has a spin with favored orientation with respect to edges 1 and 2. Thus these edges can be either blue or green, and this central spin cannot be favored with respect to any other edge. In the interesting range of 
temperatures we observe 36 $n$-stars with $4\le i+j+k\le 6$.}
\label{star}
\end{figure}
Next consider an $n$-star, which by definition is a given particle decorated by the $n$ inter-particle vector distances (edges) to its $n$ neighbors, see for example Fig. \ref{star}. Each such edge is colored according to the spin orientations. We denote by $i,j,k$ the number of red, green and blue edges such that $n=i+j+k$. It turns out that in the temperature
range of interest ($0<T<0.5$), in an overwhelming majority of $n$-stars (more than 98\%) the central particle has a spin orientation that is favorable with respect to two of its edges (this is of course the maximal value, which is favored by energy considerations). Therefore
we take a-priori $j+k\ge 2$, neglecting the very small number of instances where this does not hold. The energy of such
an $n$-star (referred to as a quasi-particle) is computed as
\begin{equation}
E_{ijk} = iE_r +jE_g+kE_b \ , \label{energies}
\end{equation}
where $k\le 2$. Note that since the energies on the RHS of Eq. (\ref{energies}) depend on temperature, so does the energy of the quasi-particles. Notwithstanding, in the interesting temperature range the temperature dependence is weak;  we take the energies of the quasi-particles as $T$-independent (we used as half the energy of a particle pair
$E_r = -0.2187,~E_g = -0.5645,~E_b= -1.5105$). The degeneracy $g_{ijk}$ of the energy level (number of quasi-particles with the same energy) is 
\begin{equation}
g_{ijk} = {2\choose k} {i+j+k-2\choose i} 2^{j+2k-2} 4^{i-k+2}\ .
\end{equation}
Since the central particle always has a spin in favorable orientation to two edges, each of these two edges must be blue or green; the first factorial is the number of possible choices of blue edges; one way if there are two (or none) of them and two ways if there is one of them. Once these
two edges are determined, there remain $i+j+k-2$ edges to choose the $i$ red from, giving rise
to the second factorial. This completes the degeneracy due to color. Next we count the number of spin orientations. There are $2-k$ green edges due to the central particle and $i$ red edges, giving us $i-k+2$ unfavorable spin orientations and $j+2k-2$ favorable ones. The number of ways to orient the unfavorable spins is $4^{i-k+2}$ and the number of ways to orient the favorable spins is $2^{j+2k-2}$. The number $4$ stems from the fact that the two favored orientations occupy an angular sector of $2\times 60^o$, leaving us with 4 sectors of $60^o$ for the unfavorable orientations. The fact that the central particle can emanate at most two favored
edges means that we have a constraint $\sum_{ijk} (j+2k)c_{ijk}\le 4$, where $c_{ijk}$ is the mol-fraction of quasi-particles having $i$, $j$ and $k$ edges of the right color. In practice, as mentioned above,
the inequality can be swapped with an equality 
\begin{equation}
\sum_{ijk} (j+2k)c_{ijk}= 4 \ . \label{constraint}
\end{equation}
\begin{figure}
\hskip -1 cm
\epsfig{width=.45\textwidth,file=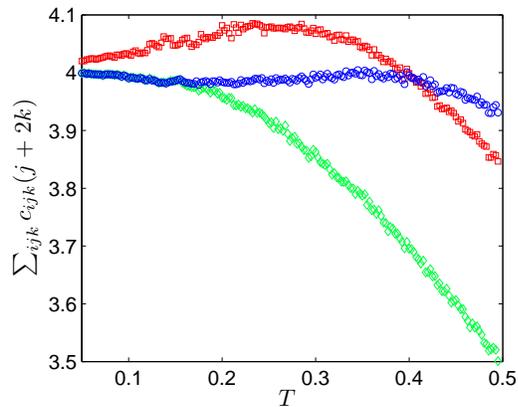}
\caption{(Color online). Direct numerical simulation of the constraint (\ref{constraint}).}
\label{Figconstraint}
\end{figure}
To satisfy this constraint exactly we need to consider the temperature dependence of the 
spin-fluctuation sectors, since when these change, so do the assignments of $(ijk)$. In Fig.
\ref{Figconstraint} we show the LHS of Eq. \ref{constraint} for two fixed spin-fluctuation sectors
(upper, red squares, $69^o$, lower, green diamonds, $50^o$, and middle, blue circles, variable sector of width $w=(68.5T+41.2)^o$). In all cases the sum
was measured via Monte Carlo simulations in the range $0.05<T<0.5$. The quality of the constraint using the variable spin-fluctuation sector is obvious. The decrease at high
temperatures is due to the increased fluctuations in  the spin orientations and in the energies
of the quasi-particles, inducing changes in the
degeneracies and in the $(ijk)$ assignments.  We thus use this temperature dependent width of
the spin-fluctuations to assign the quasi-particle index $(ijk)$ in all our simulations.

\begin{figure}
\hskip -0.8  cm
\epsfig{width=.45\textwidth,file=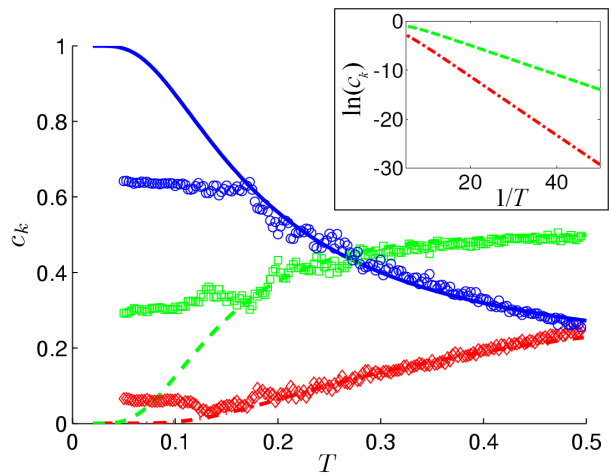}
\caption{(Color online). Comparison of the direct numerical simulations to the theoretical prediction.
Shown are the mol-fractions $c_k$ with $k=0$ (dashed-dotted red line), $k=1$ (green dashed line)
and $k=2$ (continuous blue line). One sees the point of departure of the direct numerical
simulations from equilibrium (the point of jamming) which is estimated to be about $T=0.17$. Inset: same concentrations on logarithmic scale.}
\label{results}
\end{figure}

Now write the partition function of the system:
\begin{equation}
Z(T,\lambda(T)) \equiv \sum_{ijk} g_{ijk} e^{-\beta E_{ijk}} e^{-\lambda(j+2k)} \ . \label{PF}
\end{equation}
The Lagrange multiplier $\lambda$ is introduced to insure that the constraint (\ref{constraint}) is satisfied. In terms of the
partition function the mol-fraction of quasi-particles is 
\begin{equation}
c_{ijk} =\frac{g_{ijk} e^{-\beta E_{ijk}} e^{-\lambda(j+2k)}}{ Z(T,\lambda(T))} \ .  \label{mol-fractions}
\end{equation}
Substituting Eq. (\ref{mol-fractions}) in Eq. (\ref{constraint}) we compute $\lambda(T)$ for each temperature,
and then compute the mol-fraction $c_{ijk}$. For presentation and comparison
with numerical simulations it is advantageous to bunch groups of $c_{ijk}$ together. One bunching is in the three groups obtained with $k=0,1,2$. In Fig. \ref{results} we present a comparison
of the theory to the simulation for the mol-fractions of quasi-particles with  $k=0,1,2$. We note
that the agreement is excellent down to $T\approx 0.17$ where the simulation gets jammed. This
observation is in agreement with \cite{06ST} who estimated the glass transition temperature
to be about 0.15 on the basis of the ``divergence" of relaxation times. We note that the statistical
mechanics predicts the precise spin statistics of the glassy jammed state, since the mol-fractions freeze at a ``fictive temperature" $T\approx 0.17$ that can be read directly from 
Fig. \ref{results}. We do not need to measure relaxation times to see where the system falls out of
equilibrium; it is obvious directly from Fig. \ref{results}.

In addition, the statistical mechanics predicts two ``transitions" when the mol-fractions
of quasi-species with $k=0$ and then with $k=1$ become small. The glass transition
(jamming) occurs visibly when the mol-fraction $c_{k=0}$ becomes small. 
We read a second transition when $c_{k=1}$ becomes exponentially small. This implies that the ground state consists solely of $k=2$ quasi-particles, in perfect agreement with the existence of the a crystalline ground state with anti-ferromagnetic order. Thus the offered statistical mechanics explains very well
the phenomenology of this system. Note that both these "transitions" refer to finite systems, where sufficiently small concentrations mean effectively zero concentration.

The greatest challenge for the statistical mechanics is whether it can also predict the measured
relaxation times. Jamming is caused by the rapid reduction in the mol-fraction of some spin configurations, leading to a loss of ergodicity. To see this consider the quasi-particles with $k=0$. These are the highest in energy and accordingly their  mol-fraction goes to zero first when the temperature cools down.  Using then the mol-fraction $c_{k=0}$ in comparison with the area per particle $a=A/N$, we form a length scale $\xi$ according to
\begin{equation}
\xi(T) \equiv \sqrt{a/c_{k=0}(T)} \ . \label{xi}
\end{equation}
When $c_{k=0}\!\to\! 0$ the length scale $\xi\!\to\! \infty$, defining regions of increasing size that are jammed and therefore contributing to an increasing relaxation time. Spin relaxations are dominated by
correlated stringy (1-dimensional) chains, and we estimate the number of quasi-particles involved,  $N^*$, as $N^*(T) \approx  \xi(T)$.
The relaxation time is determined by the free energy barrier, and denoting by $\mu$ the chemical
potential per quasi-particle we write \cite{06ABIMPS,04BB}
\begin{equation}
\tau_\alpha \!=\!\tau_0\! \exp\left[\frac{N^*(T)\mu}{T}\right] \!=\!\tau_0 \exp\left[\frac{\mu\sqrt{a}}{T \sqrt{c_{k=0}(T)} }\right], \label{tau}
\end{equation}
\begin{figure}
\hskip -.90 cm
\epsfig{width=.45\textwidth,file=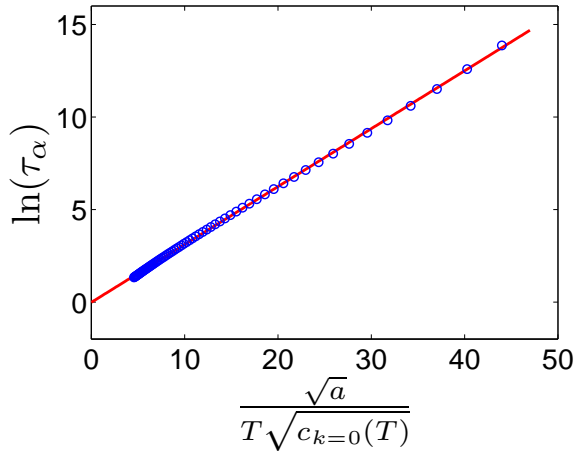}
\caption{(Color online). Comparison of the measured spin rotational relaxation times $\tau_\alpha$ (points) with the relaxation time predicted by Eq. (\ref{tau}) (continues line).} 
\label{relaxation}
\end{figure}
Note that this prediction differs essentially from the Adam-Gibbs formula \cite{AG} in the sense that
it does not predict a divergent $\tau_\alpha$ at any finite temperature, but rather an
enormous increase in $\tau_\alpha$ when  $c_{g=0}\to 0$ exponentially fast. The statistical
mechanics does not allow  $c_{g=0}= 0$. Of course, in any  finite system ``exponentially small" mol-fractions can be actually zero, and the relaxation time can be effectively infinite. The theory
does not recognize however a sharp transition in the thermodynamic limit.

The comparison of the prediction (\ref{tau}) to the measured values of the relaxation times
is shown in Fig. \ref{relaxation}. We note both the excellent agreement and the fact that $\tau_0$
is of the order of unity, as expected in the limit $T\to \infty$ where the relaxation time should
be the particle vibration time.
We again draw the attention of the reader to Fig. 3 in \cite{06ABIMPS} where a fit to the relaxation time is achieved, using similar ideas. In the problem there the relaxation was configurational
rather than via an internal variable as here, and typically relaxation events spanned  2-dimensional correlated domains, making the free energy barrier proportional to $\xi^2$. The equal usefulness of the ideas used, with the only change in the identification of the quasi-species, their degeneracy and the constraints on the statistical mechanics gives us hope that the approach is quite general and can be applied to glass forming systems of very different nature. Whether or not such a computer-assisted statistical mechanics can be applied to 3-dimensional
glass formers is a question that must await future research. 

This work had been supported in part by the German-Israeli Foundation, the Minerva Foundation, Munich, Germany, and the Israel Science Foundation.


\begin{thebibliography}{99}

\bibitem{03RHGF}
See for example Ya. E. Ryabov, Y. Hayashi, A. Gutina and Y. Feldman, Phys.Rev. B {\bf 67}, 
132202 (2003) and references therein.

\bibitem{06ST}
H. Shintani and H. Tanaka, Nature Physics {\bf 2}, 200 (2006).

\bibitem{89DAY}
D. Deng, A.S. Argon and S. Yip, Philos. Trans. R. Soc. London Se

\bibitem{99PH}
D.N. Perera and P. Harrowell, Phys. Rev. E {\bf 59}, 5721 (1999) and references therein.

\bibitem{06ABIMPS}
E. Aharonov, E. Bouchbinder, H.G.E. Hentschel, V. Ilyin, N. Makedonska, I. Procaccia and N. Schupper, Europhys. Lett. {\bf 77}, 56002 (2007).

\bibitem{07HIMPS}
H.G.E. Hentschel, V. Ilyin, N. Makedonska, I. Procaccia and N. Schupper, Phys. Rev.  E (2007).

\bibitem{91AT}
M.P.Allen and D.J. Tieldesley, {\em Computer Simulations of Liquids}, (Clarendon Press, Oxford (1991)).

\bibitem{AG}
G. Adam and J.H. Gibbs, J. Chem. Phys.  {\bf 43}, 139 (1965).

\bibitem{04BB}
J.P. Bouchaud and G. Biroli, J. Chem. Phys. {\bf 121}. 7347 (2004).


\end{thebibliography}
\end{document}